\documentstyle[11pt,aasms4]{article}

\def \N7027{NGC 7027}

\begin{document}

\title{Discovery of Extended X-ray Emission from the Planetary Nebula \N7027\
by the Chandra X-ray Observatory }

\author{Joel H. Kastner \\
\small
Chester F. Carlson Center for Imaging Science, Rochester Institute of 
Technology, 54 Lomb Memorial Dr., Rochester, NY 14623; jhk@cis.rit.edu \\
\normalsize
Saeqa D. Vrtilek \\
\small
Harvard-Smithsonian Center for Astrophysics, Cambridge, MA
02138; saku@cfa.harvard.edu \\
\normalsize
Noam Soker \\
\small
Department of Physics, University of Haifa at Oranim, Oranim, Tivon 36006, 
ISRAEL; soker@physics.technion.ac.il 
}

\begin{abstract}
We report the discovery of X-ray emission from NGC 7027, a
prototypical object for the study of the formation and
evolution of planetary nebulae (PNs). Observations with the
Advanced CCD Imaging Spectrometer (ACIS) aboard the Chandra
X-ray Observatory show that the X-ray emission from NGC 7027
is extended and is bipolar in morphology. The ACIS spectrum
displays strong emission from highly ionized Ne and weaker
emission features which we attribute to O, Mg, and Si. Model
fits to this spectrum suggest a characteristic temperature
$T_x \sim 3\times10^6$ K and an intrinsic (unabsorbed) X-ray
luminosity of $L_x \sim 1.3\times10^{32}$ ergs s$^{-1}$. The intranebular
absorption of X-ray emission is highly nonuniform, but the
modeling indicates an average column density $N_H
\sim 6\times10^{21}$ cm$^{-2}$, consistent with previous
measurements of relatively large visual extinction within
the nebula. We suggest that the X-ray emission from NGC 7027
is or was generated by a hitherto undetected fast wind from
the central star of NGC 7027, or from a companion to this
star. Chandra's detection of extended, high-temperature X-ray
emission from BD +30$^\circ$ 3639, NGC 6543, and now NGC 7027 suggests
that such emission is a common feature of young
planetary nebulae.
\end{abstract}

\keywords{stars: mass loss --- stars: winds, outflows --- 
planetary nebulae: individual (NGC 7027) --- X-rays: ISM}

\section{Introduction}

The shaping of planetary nebulae (PNs) is a topic of
considerable contemporary interest in astronomy (Kastner, Soker,
\& Rappaport 2000a). It has long been understood that a PN is the
ejected envelope of an expired red giant star, which has been
subsequently ionized and accelerated by the combination of
UV radiation and fast winds from the emerging white dwarf
that was the core of the former star. However, despite the
appeal and widespread acceptance 
of the so-called ``interacting winds'' model of PN evolution
(Kwok, Purton, \& Fitzgerald 1978), crucial
details of the morphologies and kinematics of many PNs do not appear to be
explained by such a model (Frank 2000; Soker 2000). 

The early stages of evolution of PNs (and/or proto-PNs) appear to
hold the key to an understanding of the mechanisms
ultimately responsible for shaping these objects (e.g., Sahai \&
Trauger 1998).  The nearby (distance $880\pm150$ pc; Masson
1989), well-studied NGC 7027 represents a
particularly intriguing and important object in this
regard. It is evidently a young PN (dynamical age $\sim600$ yr;
Masson 1989) and displays a remarkably complex
morphology. Optical imaging by the Hubble
Space Telescope (Bond et al.\ 1997; Ciardullo et al.\ 1999)
reveals a bright, compact ($\sim 5''$ radius) core region
encircled by concentric rings of 
reflection nebulosity that extend to $15''$ in radius (Fig.\ 1a). 
The core region clearly suffers a large
degree of spatially irregular (clumpy) extinction in the
optical whereas, in the near-infrared (Kastner et al.\
1994; Latter et al. 2000, hereafter L2000) 
and radio (Masson 1989), the ionized
region is revealed to be an elliptical shell. Surrounding this
shell, but largely interior to the system of concentric
rings, is a photodissociation region with a remarkable
cloverleaf or double-ring morphology that is best seen in
near-infrared H$_2$ and PAH emission (Graham et al.\ 1993;
Kastner et al.\ 1996; L2000). 

Further complicating
this picture is a system of at least two pairs of lobes
protruding out from the main photodissociation
region. L2000 speculate that these features, which can be
seen both in H$_2$ emission and in reflection (Fig.\ 1a), are formed by
fast, well-collimated outflows impinging on the relatively
slowly expanding ($V_{\rm exp} \sim 20$ km s$^{-1}$) 
shell of molecular gas surrounding the photodissociation
region. If such fast,
collimated flows are present, then wind interactions and/or 
magnetic fields may be important in shaping this young PN (Blackman,
Frank, \& Welch 2000).   

To investigate this possibility, we used the 
Chandra X-ray Observatory to observe NGC 7027, with the
goal of detecting extended, high-temperature gas within this
PN. According to theory (e.g., Mellema \& Frank 1995; Soker
1994), very hot ($>10^6$ K) gas is likely to be present if interacting
winds or magnetic fields play an important role in sculpting
a PN. Chandra imaging has already demonstrated the
presence of an asymmetric ``bubble'' of hot
($\sim3\times10^6$ K) gas in BD +30$^\circ$3639 
(Kastner et al.\ 2000b, hereafter KSVD) and an elongated
``bubble'' of X-ray emitting plasma in NGC 6543 (Chu et
al. 2001). Whereas X-rays were detected from both BD
+30$^\circ$3639 and NGC 6543 prior to the Chandra
observations of these nebulae, NGC 7027 had not been
detected previously in X-ray 
emission. Given the many similarities between BD
+30$^\circ$3639 and NGC 7027, however, we anticipated that the
latter also should be an X-ray source. 

\section{Observations and Data Reduction}

Chandra observed NGC 7027, with ACIS as the focal
plane instrument, on 2000 June 1.  The duration of the
observation was 18.2 ks. The Science Instrument Module was
translated and the telescope was pointed such that the
telescope boresight was positioned near the center of the
spectroscopy CCD array (ACIS-S) and the coordinates of the
PN fell on the central back-illuminated CCD (device S3).
The ACIS-S3 pixel size is $0.49''$, similar to the
spatial resolution of Chandra's High Resolution Mirror Assembly. 
The Chandra X-ray Center
(CXC) carried out standard pipeline processing on the raw
ACIS event data, producing an aspect-corrected,
bias-subtracted, graded, energy-calibrated event list,
limited to grade 02346 events (ASCA system). From this list,
we constructed a broad-band (0.3--10.0 keV) image. To
improve its signal to noise ratio, the image was convolved
with a Gaussian function with full-width at half maximum of
2.0 pixels, such that the resultant image (Fig. 1b) has an
effective resolution of $\sim1''$. {\it The image in Fig.\ 1b
represents the first detection of X-rays from this
well-studied PN.} The J2000 coordinates of the center of
the X-ray nebula (as determined from the CXC-processed event
list), RA $=$ 21$^h$07$^m$01.75$^s$, dec $=$
+42$^\circ$14$'$10.0$''$ ($\pm$0.5$''$), are the same, to
within the uncertainties, as the
coordinates of the optical nebula (as listed in the
SIMBAD\footnote{http://simbad.u-strasbg.fr/Simbad}
database). 

We also extracted the ACIS pulse height spectrum of NGC 7027. To
do so, we used CXC software to construct a histogram
of pulse heights for events contained within a circle of
radius 20 pixels ($\sim10''$), an aperture
judged to include all of the X-ray flux from
NGC 7027. The broad-band (0.2--3 keV) ACIS-S3 count rate
within this aperture was 0.014 counts sec$^{-1}$. The
background count rate in a nearby, off-source region of equivalent
area was $\sim0.001$ counts sec$^{-1}$ (0.2--3 keV). The ACIS-S3
count rate is consistent with the non-detection of NGC 7027
by ROSAT All-Sky Survey (RASS) observations in that, adopting
a plausible X-ray emission model (\S 3.2), the ACIS count
rate implies a ROSAT
Position-Sensitive Proportional Counter (PSPC) count rate of
$\sim2\times10^{-3}$ counts sec$^{-1}$, well below the typical
sensitivity limits of both the bright and faint source
catalogs (Voges et al.\ 1999, 2000).

\section{Discussion}

\subsection{X-ray Image}

The Chandra image of NGC 7027 reveals that the nebula
is clearly extended in X-rays, and
shows a distinct butterfly morphology (Fig.\ 1b). While this
morphology differs sharply from that of the
clumpy, more or less elliptical nebula seen in the
optical, the emitting region distributions correspond
in several key respects (Fig.\ 1). First, the narrow ``waist''
just southeast of the center of the X-ray nebula --- which appears to 
divide the nebula into a bright, northwest lobe and a fainter, southeast
lobe -- corresponds to a conspicuous ``dark lane'' located
just southeast of the
geometric center of the optical nebula. The optical dark lane marks
the equatorial plane of the system or, perhaps, a ring of neutral
material at high latitude (L2000) 
that is seen in absorption against
the optical nebula. This ``sense of perspective'' is
confirmed by the kinematics of molecular hydrogen emission from
NGC 7027, which indicate that the NW side of the nebula is
pointed toward the observer and the SE side is pointed away
(L2000). Second, the brighter (NW) 
X-ray lobe is located on
the side of the optical nebula that is
closer to the observer. Third, the X-ray brightness peak located
along the western edge of the NW lobe corresponds to the
brightness peak of the optical nebulosity. Finally, 
the principal direction of alignment of the X-ray emission (NW-SE)
follows that of projectile-like protrusions 
that appear in scattered light in the HST/WFPC2 color composite
(Fig.\ 1a). 

To illustrate better this last correspondence, we show in
Fig.\ 2 an overlay of contours of X-ray emission on a
near-infrared (2.12 $\mu$m) H$_2$ $+$ continuum image
obtained by HST/NICMOS (L2000). The images have been
registed such that the compact source of X-ray emission near
the center of the Chandra image coincides with the position
of the central star (see below).  This registration requires
the application of offsets of $+1.5''$, $+0.5''$ (in RA,
Decl.) to the Chandra image, which are well within the
absolute pointing uncertainties of Chandra.  The overlay
demonstrates that the X-ray emission is largely contained
within the central, elliptical shell of bright nebulosity
seen in the near-infrared. However, the brightest X-ray
emission is detected along those directions extending from
the central star toward the outermost H$_2$ filaments,
especially along the direction toward the H$_2$ feature that
extends 10$''$ to the northwest of the central star (axis 1
in Fig.\ 3 of L2000).  

The image overlay in Fig.\ 2 indicates the possible presence
of X-ray emission from the vicinity of the central star.
However, the image alignment in Fig.\ 2, while suggestive,
is not unique and cannot be used to conclude that the
central star is an X-ray source. Such emission, if present,
is not nearly so prominent as in the case of the central
star of NGC 6543 (Chu et al.\ 2001).

\subsection{X-ray Spectroscopy}

The Chandra pulse height spectrum of NGC 7027 shows that
almost all detected photons have energies between $\sim0.2$ keV 
and $\sim 2.5$ keV (Fig.\ 3). The spectrum, which peaks at
$\sim1$ keV, is evidently somewhat
harder than that of BD $+30^\circ$ 3639, which peaks at
$\sim 0.5$ keV (KSVD). However, the spectra of both
young PNs share a prominent feature
at $\sim0.9$ keV, which is likely due to a blend of Ne lines. 
The ACIS spectrum of NGC 7027 also displays weak features
at $\sim0.6$ keV, $\sim1.3$ keV and $\sim1.8$ keV, which we
tentatively attribute to emission lines of O, Mg, and Si, respectively. 

Guided by these identifications, we performed fits of a
variable-abundance plasma emission model (the VMEKAL model)
using the CXC's Sherpa software (v1.1). The results, though
highly uncertain due to the relatively small number of
counts ($\sim250$) in the spectrum, indicate an approximate
emitting region temperature $T_x \sim 3\times10^6$ K and
volume emitting measure $\sim 2\times10^{54}$ cm$^{-3}$. The
temperature is reasonably well constrained by the shape of
the Ne feature, which suggests that emission from the Ne IX complex at 13.5
\AA\ (0.89 keV) is stronger than that from the Ne X line at
12.1 \AA (1.0 keV). The
fit suggests the X-ray emitting region has near-solar
abundances of O and Ne, and overabundances of He, C, N, Mg,
and Si. As in the case of BD $+30^\circ$ 3639 (KSVD), no Fe emission
is evident in the ACIS spectrum of NGC 7027.

The intervening absorbing column derived from the fitting is $N_H \sim
6 \times 10^{21}$ cm$^{-2}$, which is in good
agreement with measurements 
of visual extinction within the nebula (e.g., $A_V = 2.97$ mag;
Robberto et al. 1993). It is apparent, however, that the
absorption is highly nonuniform. In Fig.\ 4 we present
spectra extracted for $5''\times5''$ square regions
encompassing the NW and SE X-ray lobes of the
nebula. The spectrum of the fainter SE lobe is evidently
harder than that of its brighter NW counterpart, suggesting
considerably larger intervening absorption toward the SE lobe. 
Thus, the above result for $N_H$ represents the average
absorbing column toward the dominant emission source, i.e.,
the NW lobe. Adopting the model results for mean $N_H$ and
for $T_x$, we derive a total observed flux of $F_x = 3.1 \times
10^{-14}$ ergs cm$^{-2}$ s$^{-1}$ and total unabsorbed
(intrinsic) source luminosity of $L_x = 1.3 \times
10^{32}$ ergs s$^{-1}$. 

\section{Conclusions}

The detection by Chandra of extended, high-temperature X-ray
emission in the central regions of 
BD +30$^\circ$ 3639 (KSVD), NGC 6543 (Chu et
al.\ 2001), and now NGC 7027 suggests
that many or even most young
planetary nebulae may harbor very high temperature inner
regions. Such emitting regions, including that in NGC 7027,
likely have escaped 
detection by previous X-ray telescopes due to a combination
of their compact source sizes and obscuration along
our lines of sight.
Indeed, the values of $T_x$, emission measure, and
$L_x$ derived for NGC 7027 (\S 3.2) are very similar to
those found for BD $+30^\circ$ 
3639 (KSVD). Given this similarity, and the similar apparent X-ray emitting 
volumes of the two PNs, it is apparent that the large difference in
their Chandra/ACIS-S count rates is 
due primarily to the larger absorbing column
characterizing the emission from NGC 7027. 
Furthermore, the correspondence between the optical and X-ray morphologies 
of NGC 7027 (Fig.\ 1), and the sharp
differences between the surface brightnesses and
spectral energy distributions of the two X-ray lobes (Fig.\ 4), 
strongly suggests that the ``patchy'' X-ray
appearance of NGC 7027 is determined in large part by foreground
extinction in the nebula itself. 

We conclude, therefore, that the same mechanism is
reponsible for the X-ray emission from both NGC 7027 and BD
$+30^\circ$ 3639, and that the differences between their
X-ray spectra and morphologies are largely a result of
viewing angle. That is, the two nebulae have very similar
intrinsic structures --- prolate ellipsoidal shells with
multiple protrusions along specific directions at
high latitude --- but BD $+30^\circ$ 3639 is viewed more
nearly pole-on than NGC 7027.  This interpretation is
consistent with the measurement of large molecular outflow
velocities in BD $+30^\circ$ 3639 (Bachiller et al.\ 2000)
and with the smaller visual extinction measured toward its
central star ($A_V \sim 0.75$ mag; Leuenhagen, Hamann, \& Jeffery
1996).

It is very likely that the X-ray emission from both PNs
originates in shocks formed
by the collision of fast outflows from the central star(s)
with slower-moving material within the elliptical shells
detected in radio, infrared, and optical imaging. A similar
mechanism likely explains the X-ray bubble that fills the
central region of NGC 6543 (Chu et al.\ 2001).
Both BD $+30^\circ$ 3639 and NGC 6543 
present evidence for fast winds,
with velocities of $v_f = 700$ km s$^{-1}$ (Leuenhagen et al.\ 
1996) and $v_f = 1700$ km s$^{-1}$ (Perinotto, Cerruti-Sola, \&
Lamers 1989), respectively; no such fast
wind has been detected in NGC 7027. One prediction of the
foregoing model, therefore, is that the central star of NGC
7027 --- or a companion to this central star (Soker \&
Rappaport 2000 and references therein) --- drives a
fast wind. If this fast wind were highly collimated
it would explain simultaneously the ``patchy'' X-ray
emission morphology of the nebula as well as its outer loops
of H$_2$ emission (L2000). For the X-ray emitting gas to be shocked 
to a temperature of $3\times10^6~{\rm K}$ the minimum
preshock fast wind velocity would be $\sim 400~{\rm km}~{\rm s}^{-1}$.

The electron density $n_e$ can be obtained from the emission
measure 
${\rm EM} \equiv n_e n_p V_x \simeq 2 \times 10^{54}$ cm$^{-3}$,
where $n_p$ is the proton density and $V_x$ the volume of the
X-ray emitting gas.
Assuming that the X-ray emitting gas occupies half the volume of
the inner cavity (Kastner et al.\ 2001), we estimate
$V_x=10^{50}$ cm$^3$. This suggests an average electron
density $n_e \simeq 150$ cm$^{-3}$.  
The cooling time for gas at $3\times10^6~{\rm K}$ is then given by
\begin{equation}
t_{\rm cool} \simeq 7000  
\left( \frac {n_e}{150~{\rm cm}^{-3}} \right)^{-1}
{\rm yr}, 
\end{equation}
which, given the derived electron density, is much longer
than the dynamical age of the nebula ($\sim 600~{\rm yr}$;
Masson 1989). So, if the present wind speed is well in excess of $\sim
400~{\rm km}~{\rm s}^{-1}$, as would be expected for the
present central white dwarf (see below), this would suggest 
heat conduction along magnetic field lines 
and/or mixing of the fast wind with nebular material acts to
moderate the temperature of the X-ray emitting
region (KSVD and references therein; Chu et al.\ 2001). The total mass of
hot ($T \sim 3\times10^6$ K) gas implied by our observations and modeling, 
$\sim 10^{-5} M_\odot$,
suggests a duration of only $\sim100$ yr for the episode of
mass loss via a fast wind --- less than the
dynamical age of the nebula --- assuming a mass
loss rate typical of the central stars of young PNs (i.e.,
$\sim 10^{-7} M_\odot ~{\rm yr}^{-1}$) and that the X-ray-emitting gas
is dominated by fast wind (rather than nebular) material.

The presence of a luminous source of
X-rays within NGC 7027 raises the intriguing possibility
that this emission is ultimately responsible for the
excitation of its infrared H$_2$ line emission (e.g., Gredel
\& Dalgarno 1995) and, perhaps,
for the presence of certain key species in its molecular
envelope (e.g., HCO$^+$; Deguchi et al. 1990). In a
subsequent paper (Kastner et al.\ 2001), we further 
pursue these and other suggestions raised in this paper and
in KSVD.

\acknowledgements{The authors thank the referee, You-Hua
Chu, for incisive, illuminating comments on this paper.
J.H.K. acknowledges support for this research
provided by NASA/Chandra grant GO0--1067X to RIT. N.S. thanks the US-Israel
Binational Science Foundation. S.D.V. acknowledges support
from NASA grant NAG5--6711.}

\newpage

\subsection*{Figure Captions}

\begin{description}

\item[Figure 1.] Comparison of {\it a)} Hubble Space
Telescope (HST) optical (Ciardullo et al.\
1999) and {\it b)} Chandra X-ray Observatory
images of NGC 7027. The HST color composite was generated
from images obtained through filters F555W (5550 \AA) and
F814W (8140 \AA).

\item[Figure 2.] Overlay of Chandra X-ray Observatory
image of NGC 7027 (contours; levels at 3, 5, 7, and 10
counts per pixel integrated over the 18.2 ksec observation) 
on a Hubble Space Telescope near-infrared image,
obtained at 2.12 $\mu$m (Latter et al.\ 2000). 
The HST image is dominated by H$_2$
rovibrational emission, but also includes contributions from
continuum and He I emission.

\item[Figure 3.] {\it Top:} Chandra/ACIS spectrum of NGC
7027 (solid histogram), with ``best-fit'' model overlaid
(dotted curve). The spectrum is
presented for a bin size of 52 eV. {\it Bottom:} Residuals
of the fit, in units of the measurement uncertainty for each
spectral bin.

\item[Figure 4.] Comparison of spectra extracted for NW and
SE X-ray lobes of NGC 7027. Both spectra are
presented for a bin size of 52 eV.

\end{description}

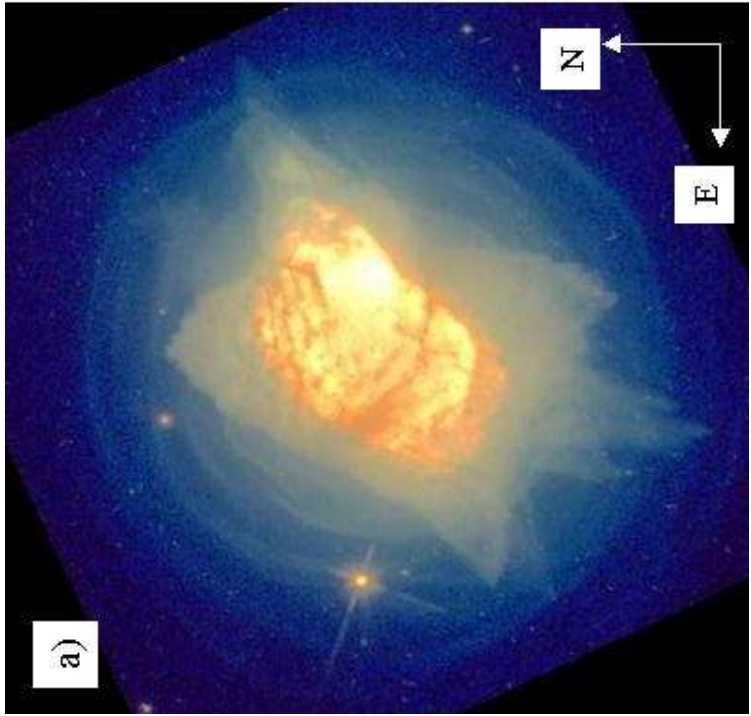
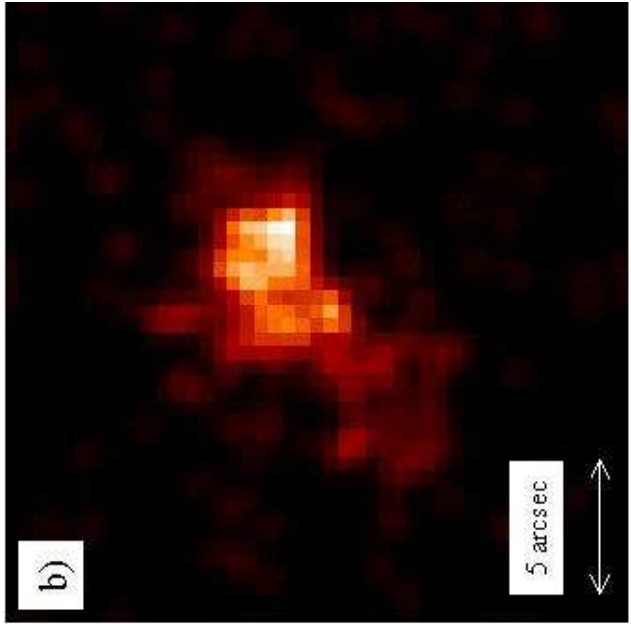
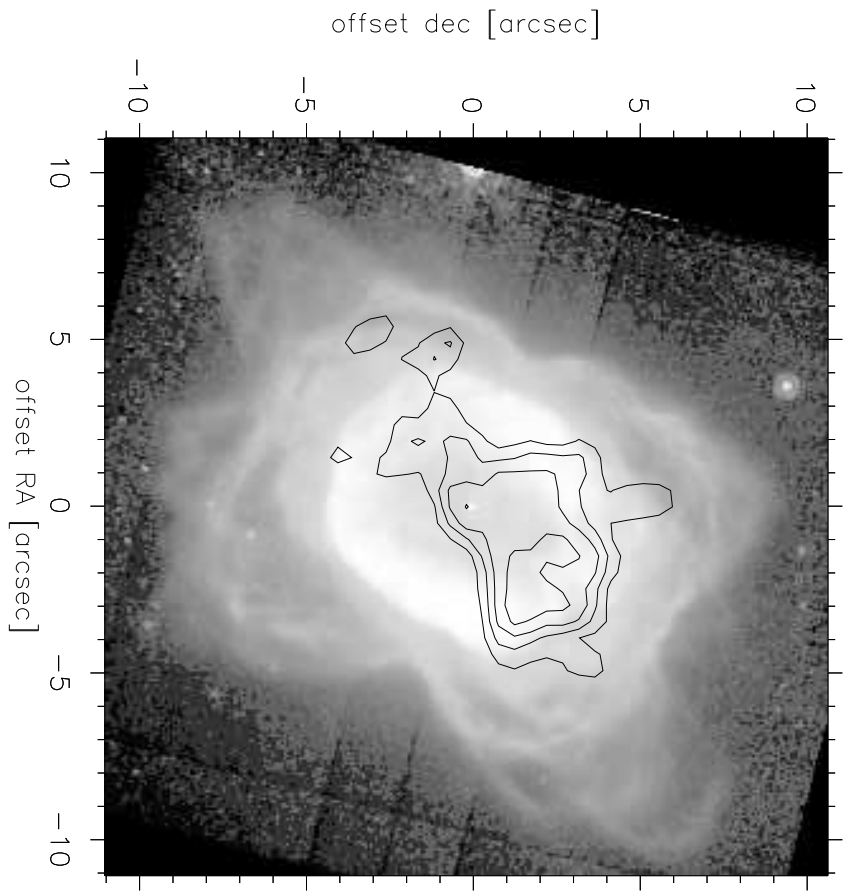
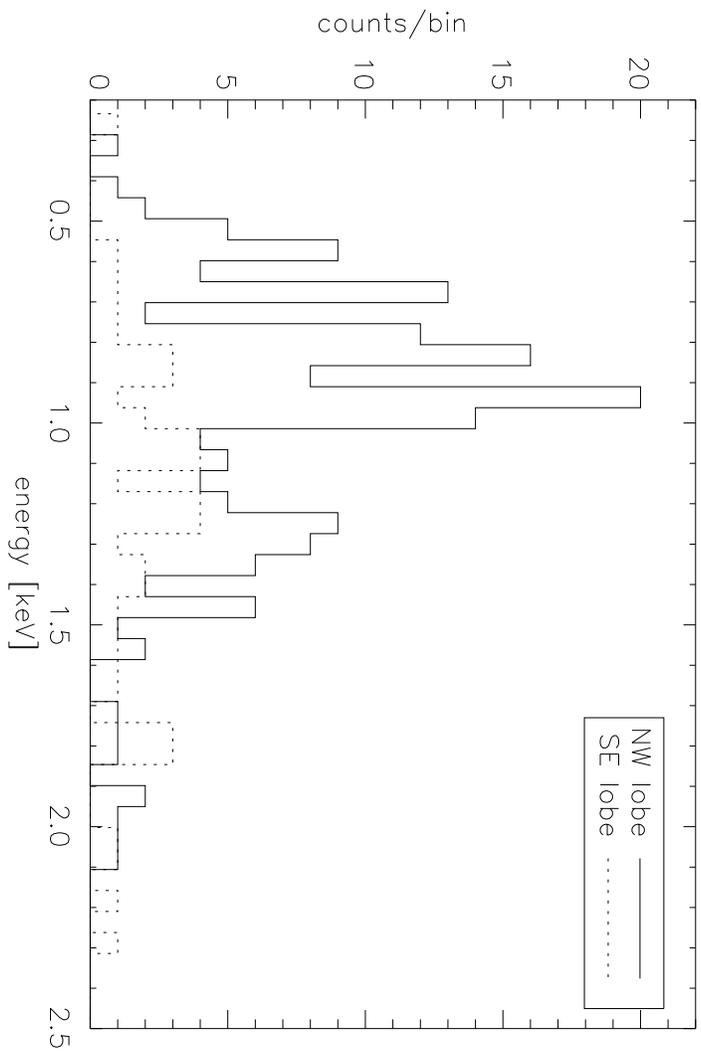
\end{document}